\documentclass{aa}
\usepackage[varg]{txfonts}
\usepackage{lipsum}
\usepackage{graphicx}
\usepackage{natbib}
\bibpunct{(}{)}{;}{a}{}{,} 

\begin{document}

\let\WriteBookmarks\relax
\def\floatpagepagefraction{1}
\def\textpagefraction{.001}

\title{Modelling the Galactic very-high-energy $\gamma$-ray source population}

\author{Constantin Steppa
\thanks{steppa@uni-potsdam.de} \and
Kathrin Egberts
\thanks{kathrin.egberts@uni-potsdam.de}}

\institute{
Universit\"at Potsdam, Institut f\"ur Physik und Astronomie, Campus
Golm, Haus 28, Karl-Liebknecht-Str. 24/25, 14476 Potsdam-Golm, Germany}
\authorrunning{C. Steppa \& K. Egberts}

\abstract{The High Energy Stereoscopic System
(H.E.S.S.) Galactic plane survey (HGPS) is to date the most
comprehensive census of Galactic $\gamma$-ray sources at very high energies
(VHE; $100\,\mathrm{GeV}\leq E\leq 100\,\mathrm{TeV}$). As a consequence of the limited
sensitivity of this survey, the 78 detected $\gamma$-ray sources comprise only a small and
biased subsample of the overall population. The larger part consists of
currently unresolved sources, which contribute to large-scale diffuse emission
to a still uncertain amount. } {We study the VHE $\gamma$-ray source population in
the Milky Way. For this purpose population-synthesis models are derived based on
the distributions of source positions, extents, and luminosities.} {Several azimuth-symmetric and spiral-arm models are
compared for spatial source distribution.
The luminosity and radius function of the population are derived from the source
properties of the HGPS data set and are corrected for the sensitivity bias of
the HGPS.
Based on these models, VHE source populations are simulated and the subsets of
sources detectable according to the HGPS are compared with HGPS sources.
} {The power-law indices of luminosity and radius functions are determined to
range between $-1.6$ and $-1.9$ for luminosity and $-1.1$ and $-1.6$ for radius.
A two-arm spiral structure with central bar is discarded as spatial distribution
of VHE sources, while azimuth-symmetric distributions and a distribution
following a four-arm spiral structure without bar describe the HGPS data
reasonably well.
The total number of Galactic VHE sources is predicted to be in the range from 800
to 7000 with a total luminosity and flux of $(1.6-6.3) \cdot
10^{36}$~ph~s$^{-1}$ and $(3-15) \cdot 10^{-10}$~ph~cm$^{-2}$~s$^{-1}$,
respectively.
} {Depending on the model, the HGPS sample accounts for $(68-87)\%$ of the
emission of the population in the scanned region.
This suggests that unresolved sources represent a critical component of the
diffuse emission measurable in the HGPS.
With the foreseen jump in sensitivity of the Cherenkov Telescope Array, the
number of detectable sources is predicted to increase by a factor between 5 - 9.
}


\maketitle

\section{Introduction}
The past two decades have witnessed the birth and explosive development of teraelectronvolt
astronomy. A major breakthrough for the development of the field and especially
the Galactic very-high-energy (VHE; $100\,\mathrm{GeV}\leq E\leq
100\,\mathrm{TeV}$) $\gamma$-ray sky has been the High Energy Stereoscopic
System (H.E.S.S.) Galactic plane survey (HGPS). For 12 years H.E.S.S. has scanned the
central part of the Milky Way (extending from Galactic longitudes of $l =
250^\circ$ to $65^\circ$ and covering latitudes of $|b|\leq 3^\circ$) and
acquired a data set of nearly 2700 hours of good-quality observations
\citep{catalog}. The HGPS has revealed a plethora of $\gamma$-ray sources
\citep{catalog} and a faint component of a so-called diffuse (large-scale
unresolved)  emission \citep{diffuse}.
With the brightest and closest sources, the sample of detected $\gamma$-ray
sources represents only the tip of the iceberg of the overall population of VHE
$\gamma$-ray emitters.
A larger percentage of sources are expected to remain unresolved with the given
H.E.S.S. exposure and sensitivity due to being too faint and/or too far away to
be significantly detected, thus forming a contribution to the measured
large-scale diffuse emission. \\
Previous studies of the VHE-detected $\gamma$-ray source classes that are based
on the HGPS characterise the sample of pulsar wind nebulae (PWNe; \cite{PWNPop})
and supernova remnants (SNRs; \cite{SNRPop}), but with limited insight into their
respective population.
A characterisation of the overall population of sources can be achieved by
population synthesis with the simulation of synthetic source samples and
comparison with observations in the range of detectability of the data set.
This procedure is customarily followed for the study of object classes such as
pulsars~\citep{Gonthier2018}.
The study of a {\it generic} source population is a slightly different approach. Rather than aiming to derive properties of a specific class of objects,  this approach characterises the overall population of sources at a certain wavelength.
This procedure allows the prediction of the number of sources detectable with future
instruments (e.g. the Cherenkov Telescope Array (CTA) \citep{Funk}) and to
estimate the amount of unresolved sources that contribute to the diffuse
emission measurements.\\
In this work, we follow this latter strategy to describe the VHE $\gamma$-ray emitters
generically, and we derive luminosity as well as radius functions for this
generic VHE $\gamma$-ray source population.
A similar approach has already been applied to data from the Energetic Gamma
Ray Experiment Telescope on board the Compton Gamma Ray Observatory and data
from the Large Area Telescope on board the Fermi Gamma-ray Space Telescope
(Fermi-LAT) for the estimation of unresolved sources in the high-energy (HE;
$100\,\mathrm{MeV}\leq E\leq 100\,\mathrm{GeV}$) Galactic diffuse
$\gamma$-ray emission \citep{Strong,Fermi3rdSourceCatalog}. As already
identified by \cite{CasanovaDingus} for the case of the diffuse emission
measured by MILAGRO~\citep{MILAGRODiffuse}, compared to HE the
contribution of unresolved sources is expected to rise, and likely dominates, at
VHE.
This is also reflected in recent measurements of Galactic diffuse emission
above 1~TeV by the High-Altitude Water Cherenkov Observatory
(HAWC)~\citep{HAWCDiffuse} which, like the H.E.S.S. measurements at
1~TeV~\citep{diffuse}, overshoot predictions considerably. Only an assessment of the entire Galactic source population  can disentangle the two components of unresolved $\gamma$-ray-source emission and diffuse emission from propagating
cosmic rays and allow for the study of cosmic-ray propagation properties in
the H.E.S.S. and HAWC data sets.

\section{Construction of the model}
\label{SEC:construction}
The VHE source population model presented in this work consists of two distinct
components: the spatial distribution of sources and distribution of
source properties, that is their luminosities and radii. To determine the spatial
distribution, we tested various models based on the assumed source classes and
the Galactic structure.
We followed a data-driven approach to derive the distribution of source
properties. Alternatively, this distribution can be derived from detailed source
modelling. However, a population model based on individual source models
involves a fair amount of assumptions, for instance about source classes
contained in the population, the age of these sources, and their environmental
conditions. In contrast, for the data-driven approach we only assume
that the source sample is representative for the population in its range of
detectability and that sources are distributed according to certain spatial
models. In the following, the derivation of each component of the model is
described in detail. After assessing the spatial distribution, the derivation of the second
component is described, which is based on a combination of the spatial model
with observed quantities of the HGPS source sample, namely integrated flux above
$1\,\mathrm{TeV}$ (henceforth referred to as flux), angular extent, and
location in the sky.

\subsection{Spatial distribution}
The number of detected Galactic VHE $\gamma$-ray sources is yet too small to
determine the spatial distribution of the entire population. However, it is
possible to construct models of the spatial distribution based on few reasonable
assumptions. Since most of the known sources are associated with SNRs or PWNe, the corresponding distributions of SNRs and
pulsars can be used as templates. Their source densities $\rho$ are well
described by an azimuth-symmetric function that only depends on the
Galactocentric distance $r$ and the height over the Galactic disc $z$ as follows:
\begin{equation}
\rho(r,z) =
\left(\frac{\left(r+r_{off}\right)}{\left(R_{\odot}+r_{off}\right)}\right)^{\alpha}\,
\exp\left({-\beta\frac{\left(r-R_{\odot}\right)}{\left(R_{\odot}+r_{off}\right)}}\right)\,
\exp\left({-\frac{\left|z\right|}{z_{0}}}\right)\,,
\label{EQ:symmetric}
\end{equation}
where $R_{\odot}$ is the distance of the sun to the Galactic centre, $z_{0}$ the
scale height of the Galactic disc, shape parameter $\alpha$, and rate parameter
$\beta$. The parameter $r_{off}$ accounts for a non-zero density at $r=0$. Based
on the assumption that SNRs are the dominant class of $\gamma$-ray sources we
probe a model (mSNR) by applying Eq.~\ref{EQ:symmetric} with parameters as given
in \cite{Green2015} and \cite{Xu2005}. Likewise, we probe a model that is based
on the assumption that PWNe are the dominant class (mPWN) using parameters as
given in \cite{Yusifov2004} and \cite{Lorimer2006}.
Both parameter sets are listed in Table~\ref{TAB:symmetric}.
\begin{table}[htb]
\centering
\caption{Parameter values corresponding to Eq.~\ref{EQ:symmetric}}
\label{TAB:symmetric}
\begin{tabular}{lccccc}
\hline
Model & $R_{\odot}\,\mathrm{[kpc]}$ & $r_{off}\,\mathrm{[kpc]}$ & $\alpha$ &
$\beta$ & $z_{0}\,\mathrm{[kpc]}$ \\
\hline
mSNR & 8.5 & 0 & 1.09 & 3.87 & 0.083 \\
mPWN & 8.5 & 0.55 & 1.64 & 4.01 & 0.18 \\
\hline
\end{tabular}
\end{table}
\\While mSNR and mPWN are two examples for azimuth-symmetric
source distributions, there is good reason to assume that the spatial
distribution of $\gamma$-ray sources might deviate from this symmetry. The
progenitors of VHE sources, for instance massive stars, typically form in dense regions of gas and dust.
Therefore, the distribution of VHE sources might be affected by
the spiral structure of the Galaxy,  for instance, observed in the distribution of
interstellar matter (ISM; \cite{SteimanCameron}). Following the study of
\cite{Kissmann2015} on the impact of spiral-arm source distributions on the
Galactic cosmic-ray flux we probed three different models of a non-symmetric
source distribution. To represent a four-arm distribution, we adopted the
model by \cite{SteimanCameron}, denoted as mSp4. Compared to
Eq.~\ref{EQ:symmetric}, in this case the source density explicitly depends on the
azimuth $\phi$ and is described as follows:
\begin{equation}
\label{EQ:spiral}
\begin{aligned}
\rho(r,\phi,z) = &\sum_{i=1}^{4}A_{i}\,
exp\left(-\frac{1}{\delta^{2}}\left(\phi-\frac{\ln\left(\frac{r}{a_{i}}\right)}{\beta_{i}}\right)^{2}\right)\\
&\cdot exp\left(-\frac{\left|r-R\right|}{\sigma_{r}}\right)\,
exp\left(-\frac{z^{2}}{2\sigma_{z,2}^{2}}\right)\,.
\end{aligned}
\end{equation}
The radial dependence is defined by the scale length $\sigma_{r}$ and a local
maximum at $R$. Likewise, the azimuthal dependence is defined by the scale
length $\delta$ and two constants, $\beta_{i}$ determining the pitch angle of
the spiral and $a_{i}$ giving its orientation. Finally, the $z$-dependence is
governed by the scale height $\sigma_{z,2}$. For this model we adopted the best-fit values
corresponding to the ISM measurement traced by CII cooling lines (see
Table~\ref{TAB:CII_fit_vals}).
\begin{table*}[t]
\centering
\caption{Parameters for the description of the Galactic spiral arms. Values as
given in \cite{SteimanCameron} for their four-arm model fitted to the ISM
distribution traced by CII emission.}
\label{TAB:CII_fit_vals}
\begin{tabular}{ccccccccc}
\hline
Spiral & $\beta_{i}$ & $a_{i}$ & $R$ [kpc] &
\multicolumn{2}{c}{$\sigma_{r}$ [kpc]} & $\sigma_{z,2}$ [kpc] & $\delta$ [deg] &
$A_{i}$ \\
\cline{5-6}
Arm &  &  &  & $(r < R)$ & $(r > R)$ &  &  & \\
\hline
Sagittarius-Carina & 0.242 & 0.246 & 2.9 & 0.7 & 3.1 & 0.070 & 15 & 169
 \\
Scutum-Crux & 0.279 & 0.608 & 2.9 & 0.7 & 3.1 & 0.070 & 15 & 266 \\
Perseus & 0.249 & 0.449 & 2.9 & 0.7 & 3.1 & 0.070 & 15 & 339 \\
Norma-Cygnus & 0.240 & 0.378 & 2.9 & 0.7 & 3.1 & 0.070 & 15 & 176 \\
\hline
\end{tabular}
\end{table*}
We adopted another four-arm model with different, less pronounced arm profiles from \cite{Cordes2002}, which is based on the free electron density as traced by pulsar
dispersion measurements. We refer to this model as mFE. To calculate the source
density for this model we made use of the code provided by the
authors\footnote{The code is
available at \url{http://www.astro.cornell.edu/~cordes/NE2001/}.}. Finally, we
also probed a two-arm model with an additional central bar, whose existence in our Galaxy has been indicated by Spitzer data~\citep{Spitzer} and more recently confirmed by Gaia data~\citep{Anders2019}. This model, which we refer to as mSp2B, was adopted
from \cite{Werner2015}. It is based on the model of \cite{SteimanCameron} but
only includes the Scutum-Crux and Perseus arms. The additional component for the
Galactic bar is given by
\begin{equation}
\label{EQ:bar}
    \rho_{bar}(r,\phi,z) = 
\begin{cases}
    A_{bar}\,\exp\left(-\frac{z^2 +
    r^{2}\left(\sin(\phi)-\cos(\phi)\sin(\theta)\right)^{2}}{\sigma_{z,1}^{2}}\right),
    &\text{if } r < l_{b}\\
    0,              & \text{otherwise}
\end{cases}
\end{equation}
with its radial extent $l_{b}=3.5\,\mathrm{pc}$, rotation angle relative to
the solar-galactic centre line $\theta = 30.0\,\mathrm{deg}$ and scale
height $\sigma_{z,1} = 0.31\,\mathrm{pc}$. The normalisation is chosen such that the bar
is equally contributing to the source density as the spiral arms
($A_{bar}=364$).\\\\
For our study we simulated the source distribution by using a set $S_{sim}$ of
$\sim2\cdot10^{6}$ uniformly, randomly distributed points in a box of size
$30\,\mathrm{kpc}\times 30\,\mathrm{kpc}\times10\,\mathrm{kpc}$ yielding a
mean distance among points of $\sim35\,\mathrm{pc}$. Each point is then weighted
according to the source density given by the tested model\footnote{Owing to computational
constraints a reduced data set with $\sim2\cdot10^{5}$ points was used to probe
mFE yielding a mean distance of $\sim75\,\mathrm{pc}$.}

\subsection{Luminosity and radius distribution}
\label{SEC:lum_size}
Next to the spatial distribution, each model comprises a distribution function
for source properties, namely luminosity ($L$) and radius ($R$). Here we assume
that the variables L and R are independent and each one follows a power law such
that the joint probability density function (PDF) $P(L,R)$ can be written as
\begin{equation}
\label{EQ:LR_PDF}
P(L,R)=N\left(\frac{L}{L_{0}}\right)^{\alpha_{L}}\left(\frac{R}{R_{0}}\right)^{\alpha_{R}}
,\end{equation}
with scaling factors $L_{0}$, $R_{0}$ and a normalisation factor $N$ that
depends on the boundaries set for L and R.
The number of detected sources $N_{det}$ is related to Eq.~\ref{EQ:LR_PDF} via
\begin{equation}
\label{EQ:N_det}
N_{det}=N_{FoV}\int dL\int dR\,C(L,R)\,P(L,R)
,\end{equation}
where the observation bias inherent to the sample of detected sources is
accounted for by the correction function $C(L,R)$ and the total number of
sources in the probed field of view (FOV) is $N_{FoV}$.
In order to reconstruct the parameters of the global distribution function from
the biased sample of detected sources, we applied a likelihood maximisation as
follows. Dividing the probed $L\times R$ phase space in equally sized bins of
$0.1\times0.1$ on logarithmic scale, we derived the true number of detected
sources $N_{true, i}$ that lie within a bin $i$ of this phase space from the HGPS catalogue. The
expected number of sources $N_{pred, i}$ is approximated via
\begin{equation}
\label{EQ:N_pred}
N_{pred, i}=C(L_{i},R_{i})\,N_{FoV}\int_{L_{min, i}}^{L_{max, i}}
dL\int_{R_{min, i}}^{R_{max, i}} dR\,P(L,R)
,\end{equation}
where the correction function is only evaluated at the centre of the respective
bin. With this type of counting exercise the distribution of the true number of
detected sources per bin is expected to follow a Poissonian $P_{\lambda}(N_{true})$ with
$\lambda=N_{pred}$. Thus, the log likelihood for the maximisation is
\begin{equation}
\label{EQ:likelihood}
\log L =\sum_{i}\log\left(P_{N_{pred, i}}\left({N_{true, i}}\right)\right)
,\end{equation}
where we sum over the grid of bins. We applied this method to derive the two
power-law indices of the joint PDF $P(L,R)$\footnote{The parameter $N_{FoV}$
from Eq.~\ref{EQ:N_pred} also enters the maximisation as free parameter but
is disregarded in the following discussion as we derive the
total number of Galactic VHE sources later.}. The distinct feature of
this procedure is the recognition and inclusion of the observation bias
corresponding to the analysed sample of sources and in particular the consideration of 
its dependency on the radius of sources, which is calculated in
Section~\ref{SEC:bias}.

\subsubsection{Source selection}
Although the HGPS catalogue results from the most systematic search for VHE
$\gamma$-ray sources to date, it suffers from several deviations of its generation from a fully automated
pipeline. These deviations are a result of the large extent of many VHE $\gamma$-ray sources and their
associated complex morphologies and complicate a treatment in a
population-synthesis approach. The HGPS combines sources that are detected by a
fixed pipeline based on maps of the detection significance for a correlation
radius of $R_c = 0.1^\circ$ {\it or} $R_c = 0.2^\circ$, plus sources labelled as
external, which are detected and characterised by custom-tailored analyses \citep{catalog}.
Furthermore, the Gaussian components obtained by the automated detection
pipeline are then manually merged into $\gamma$-ray sources, resulting in a
description of their complex morphologies as a combination of various Gaussians.
These HGPS procedures render a rigorous treatment of the data in a  population
synthesis almost impossible. With the approach followed in the study presented
in this work, we limited ourselves to {\it extended} sources with {\it known distance} in
order to derive their L-R distribution. Point-like sources were excluded from the
sample because they lack the extension information that is necessary for the
determination of $P(L,R)$. The criterion of a distance estimation being
available guarantees that flux and angular size measurements can be transformed
into $L$ and $R$ values.\\
For simplification we treated extended sources as being observed as symmetric
two-dimensional Gaussians in the projected plane on the sky. The angular extent
$\sigma_{source}$ refers to the $68\,\%$ containment radius of the measured
flux. Sources with complex morphologies, for instance shell-like structures,
were treated the same way, in which the shell radius is taken as $\sigma_{source}$.
According to the extended nature of these sources the sensitivity map with
correlation radius of $R_c = 0.2^\circ$ was used to describe the correction
function that accounts for the observation bias.
Out of the 78 sources in the HGPS, 64 pass the sensitivity threshold with $R_c =
0.2^\circ$. Selecting for extended sources reduces this sample to 50 sources.
From those 50 sources only 16 ($32\%$) are firmly identified sources with
available distance estimates. From this small sample we derived the parameters of
the luminosity-radius PDF. The distributions of observable quantities, that is
flux, extent, and composition, of the sample of these 16 selected sources do not
deviate significantly from the distributions derived for the complete sample of
sources except for the missing class of binaries\footnote{All three known
binaries in the HGPS are point-like.}.
Thus, it is assumed that the small sample is representative and distance estimates are independent of
the luminosities and radii of sources. In addition, the boundaries on L and R for the
phase space that we probe are derived from this data set as well, yielding
$10^{32.1}\,\mathrm{ph\,s^{-1}} \leq L \leq 10^{34.8}\,\mathrm{ph\,s^{-1}}$ and
$10^{0.4}\,\mathrm{pc} \leq R \leq 10^{1.7}\,\mathrm{pc}$. Thus, the presented
models cover a dynamical range of almost three orders of magnitude in the luminosity
({\it cf.} with a dynamical range of three orders of magnitude chosen by
\cite{Strong} and five orders of magnitude chosen by
\cite{Fermi3rdSourceCatalog}) and one order of magnitude in size. The scale
factors in Eq.~\ref{EQ:LR_PDF} are set to $R_{0}=1\,\mathrm{pc}$ and
$L_{0}=10^{34}\,\mathrm{ph\,s^{-1}}$. As can be derived from the units, we only
considered the number of emitted photons above 1~TeV per second as proxy for the
luminosity. The luminosity function in units of erg/s can be derived by scaling
with a characteristic mean photon energy. To calculate this mean photon energy
an additional assumption about the shape of the spectral energy distribution
(SED) of sources is required. Assuming the SED follows a power law, a mean
spectral index of $-2.4$ is found for the HGPS source sample, which yields a
mean photon energy of $3.52\,\mathrm{erg\,ph^{-1}}$ in the energy range
$1\,\mathrm{TeV}$-$10\,\mathrm{TeV}$.

\subsubsection{Correction function determination}
\label{SEC:bias}
\begin{figure}[htb]
        \centering
                \includegraphics[width=.45\textwidth]{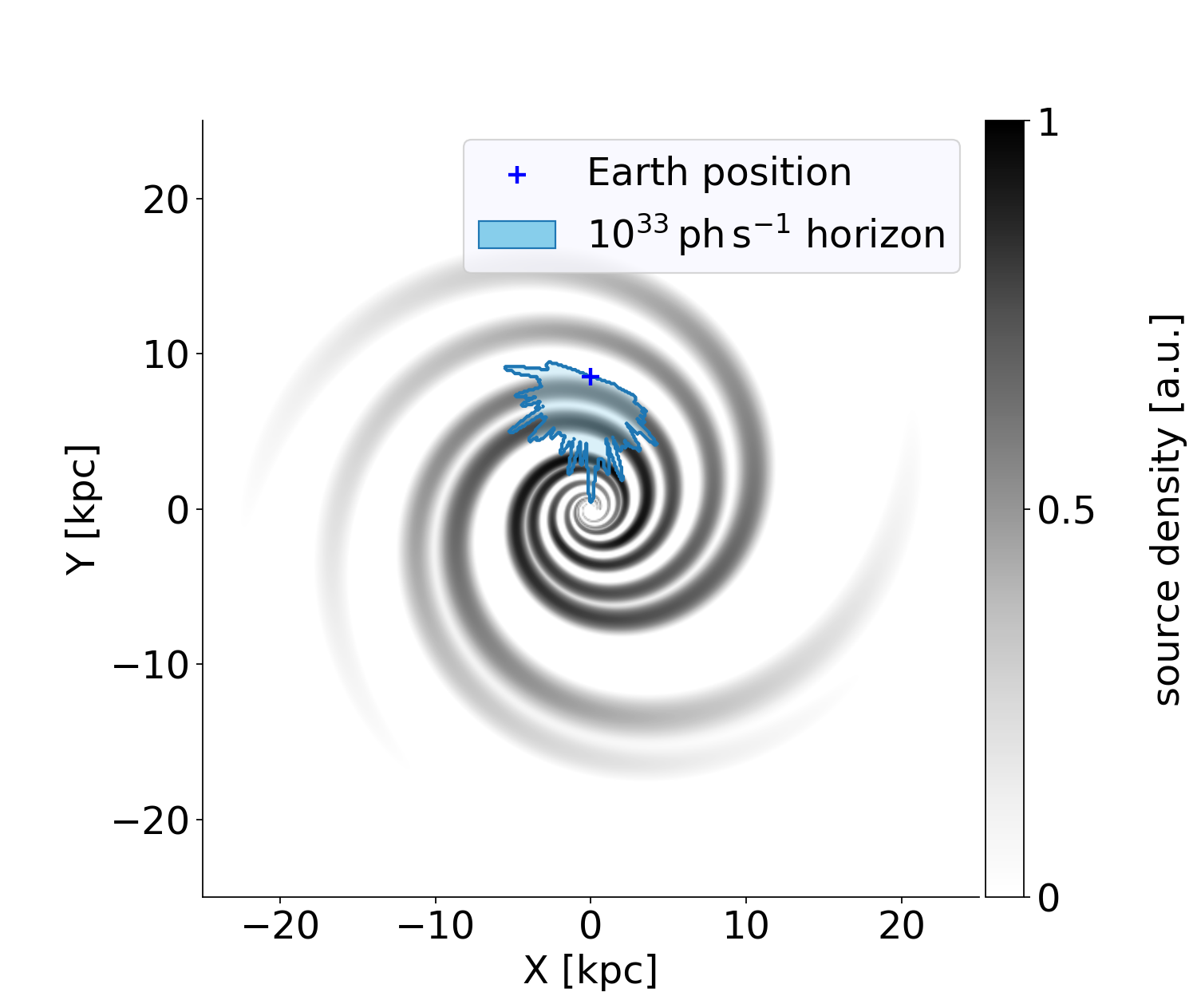}
        \caption{Distribution of VHE sources in the Milky Way following the
        four-arm spiral model by \cite{SteimanCameron} as described in
        Sec.~\ref{SEC:construction} (grey contours). Additionally, the HGPS detection
        horizon for point-like sources with a luminosity of
        $10^{33}\,\mathrm{photons\,s^{-1}}$ is plotted on top (blue contours).}
        \label{FIG:mw_model}
\end{figure}

\begin{figure}[htb]
        \centering
                \includegraphics[width=.45\textwidth]{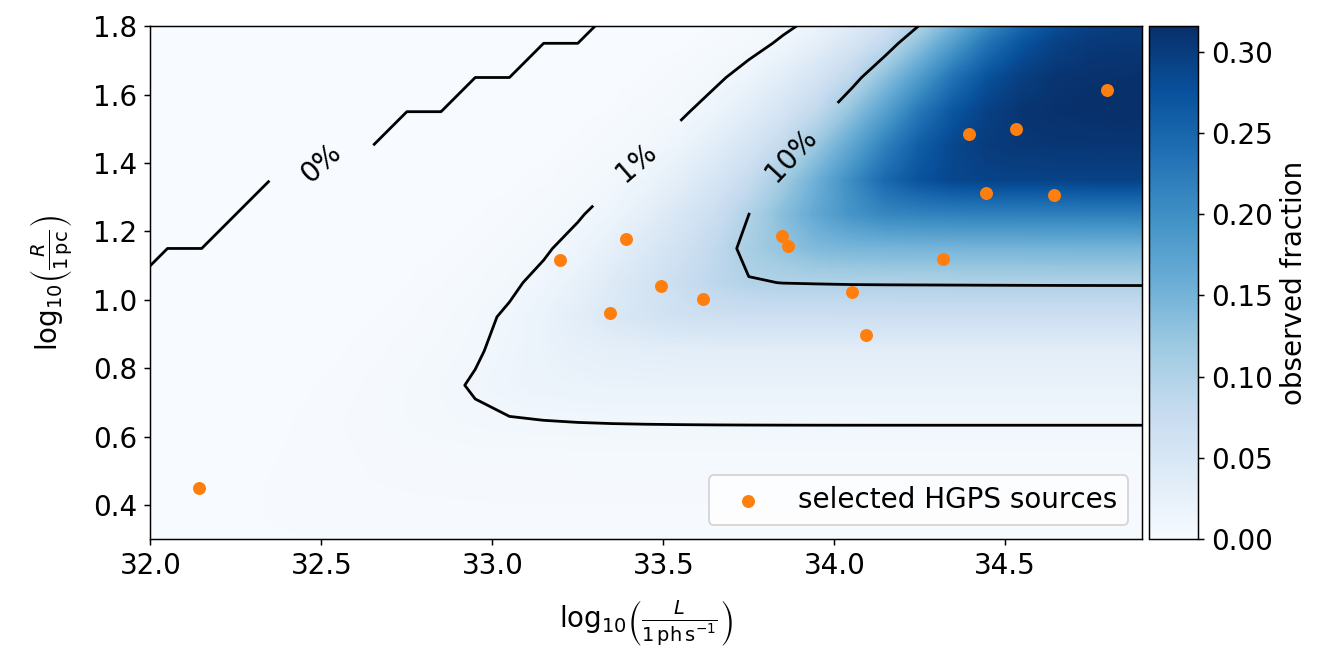}
        \caption{Estimated correction function for mSp4 showing the
        fraction of sources in the FOV of the HGPS that for a given
        luminosity and radius are detectable according to
        Eq.~\ref{EQ:adopted_sens}. The low values for small sources result 
        from the rejection of point-like sources in the analysis.
        Additionally those sources of the HGPS that fulfil the selection
        criterion are shown.}\label{FIG:bias}
\end{figure}

Derivation of the VHE $\gamma$-ray source population properties from
observational data needs to account for the strong selection bias in the
H.E.S.S. catalogue, which can be expected to distinctly shape the sample of
detected $\gamma$-ray sources and is based on the HGPS sensitivity.
The HGPS sensitivity varies strongly as a function of Galactic longitude and
latitude due to the observation pattern, which is a combination of dedicated
survey observations and additional follow-up observations and in-depth
measurements of detected sources. The inhomogeneity of the HGPS sensitivity as
a function of Galactic longitude and latitude is demonstrated in
Fig.~\ref{FIG:mw_model} through the detection horizon of point-like sources with
a luminosity of $10^{33}\,\mathrm{ph\,s^{-1}}$.\\
Besides the direction dependency, the sensitivity is also a function of the
angular extent of a source since the number of background events increases with
$\sigma_{source}$. To exceed the detection threshold of $5\sigma$ above
background, the flux of an extended source needs to be greater than
\begin{equation}
\label{EQ:scale_threshold}
    F_{min}(\sigma_{source}) = 
\begin{cases}
    F_{min,0}
    \sqrt{\frac{\sigma_{source}^{2}+\sigma_{PSF}^{2}}{\sigma_{PSF}^{2}}},
    &\leq 1^{\circ}\\
    \infty, &>1^{\circ}
\end{cases}
\end{equation}
\citep{catalog}, where $F_{min,0}$ is the point-source sensitivity,
$\sigma_{source}$ the source extent, and $\sigma_{PSF}$ the size of the H.E.S.S.
point-spread function (PSF).
In addition, the limited FOV of the instrument in combination with the
applied background subtraction technique of deriving background measurements
from within the FOV renders sources $\gtrsim 1^{\circ}$ not
detectable. For our purpose we adopted Eq.~\ref{EQ:scale_threshold} to account
for the fact that we only selected for extended sources. The HGPS does not
provide a criterion for the minimal detectable extent of a source. Therefore, we
set the threshold to the cited average value of the PSF
($\sigma_{PSF}=0.08^{\circ}$), which yields a sensitivity for extended sources that is written as\begin{equation}
\label{EQ:adopted_sens}
    F_{min}^\mathrm{extended}(\sigma_{source}) = 
\begin{cases}
        \infty, &\sigma_{source} \leq \sigma_{PSF}\\
    F_{min,0}
    \sqrt{\frac{\sigma_{source}^{2}+\sigma_{PSF}^{2}}{\sigma_{PSF}^{2}}},
    &\sigma_{PSF} < \sigma_{source} \leq 1^{\circ}\\
    \infty, &1^{\circ} < \sigma_{source}\,.
\end{cases}
\end{equation}
We define the correction function as the fraction of detectable sources, namely
sources within the adopted sensitivity range of the HGPS, in the total amount of
sources in the FoV of the HGPS. Since the sensitivity decreases for
faint or extended sources, the correction function depends on source properties, that is luminosity
and radius $C(L,R)$.\\
Under the assumption that average source properties are identical throughout the Milky Way,
we expect sources of any given properties to follow the same spatial
distribution function. To calculate the correction function, we first derive the
subset of sources which lie within the FOV of the HGPS from the set of
simulated sources $S_{FoV} \subset S_{sim}$. Assigning each source the same
luminosity $L$ and source radius $R$ we can then calculate the corresponding fluxes, angular extents, and
locations in the sky as they would be observed at Earth. Based on these
observables and the sensitivity limit (Eq.~\ref{EQ:adopted_sens}) of the HGPS,
the subset of detectable sources for the given luminosity and radius can be derived  $S_{det}(L,R) \subset S_{FoV}$. With this,
a two-dimensional correction function $C(L_{i},R_{i})$ is derived for the same
grid of source luminosities and radii mentioned above via
\begin{equation}
C(L_{i},R_{i}) =0.32\cdot\displaystyle\sum_{\mathbf{x}\in
S_{det}(L_{i},R_{i})}\rho\left(\mathbf{x}\right) /
\displaystyle\sum_{\mathbf{x}\in S_{FoV}}\rho\left(\mathbf{x}\right)\,,
\end{equation}
where $\rho$ is the source density corresponding to the assumed model. The
additional factor $0.32$ accounts for the fact that $68\,\%$ of the detected
sources are disregarded owing to missing distance estimates. 
The correction function for the spatial distribution model mSp4 is shown in Fig.~\ref{FIG:bias}
together with the distribution of HGPS sources that fulfil our selection
criterion.
Based on this estimation we show, for instance, that the distribution of
the luminosity for $L<10^{33}\,\mathrm{ph\,s^{-1}}$ cannot be well constrained by
the HGPS data. The estimation of the luminosity distribution in this regime is
affected by statistical fluctuations of the data and special care has to be
taken to explore the range of validity of the model.

\subsubsection{Monte Carlo verification}
\begin{figure}[htb]
        \centering
                \includegraphics[width=.45\textwidth]{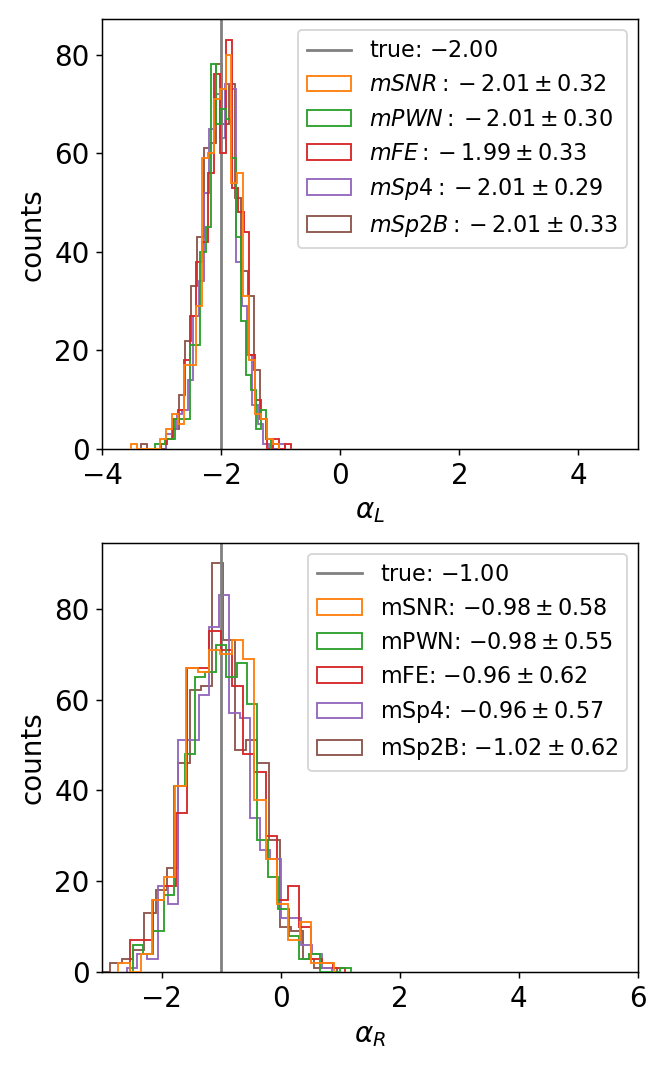}
        \caption{Distribution of reconstructed $\alpha_L$ (top) and $\alpha_R$
        (bottom) for 600 toy models using different spatial models and true values of
        $\alpha_{L,\mathrm{true}} = -2$ and $\alpha_{R,\mathrm{true}} = -1$. True
        values are denoted by the vertical line.}
        \label{FIG:MCverification}
\end{figure}

\begin{figure}[htb]
        \centering
                \includegraphics[width=.45\textwidth]{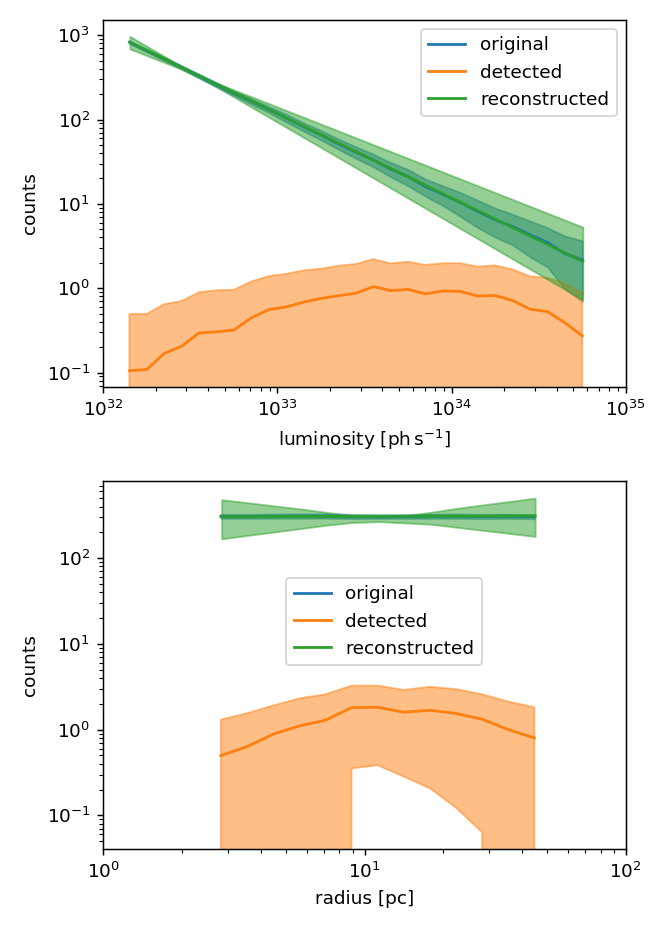}
        \caption{One-dimensional luminosity distribution (top) and radius
        distribution (bottom) for simulated populations. Original distributions are
        given in blue; distribution of detected sources in orange and the bias-corrected reconstruction of the original 
        distribution are shown in green. Details are given in the text.}
        \label{FIG:mSp4_hist}
\end{figure}

We studied the capability of reconstructing properties of the parent population of the
method presented in this work by the means of Monte Carlo simulations.
For this purpose, we simulated source populations with a set of
luminosity and radius functions for all spatial models discussed previously.
For each simulated population, $N$ spatial coordinates $\mathbf{x}$ were
randomly drawn following the distribution defined by Eq.~\ref{EQ:symmetric},
Eq.~\ref{EQ:spiral}, or Eq.~\ref{EQ:spiral} + Eq.~\ref{EQ:bar}, according to the spatial model, with
model specific parameters, together with $N$ random samples of
$L$ and $R$ following Eq.~\ref{EQ:LR_PDF} with the given parameters.
For each combination of a luminosity and radius function and a source
distribution, the subsample of detected sources was determined according
to Eq.~\ref{EQ:adopted_sens}.
As for the 50 extended sources detected with the HGPS the statistics of the
simulated source population was adjusted to yield on average the same number of
detected sources. To account for the fact that only a fraction of
$32\,\%$ of the HGPS sources comes with a distance estimation, all but 16 of the
detected simulated sources were randomly discarded. Each data set was then
reconstructed using the machinery discussed before and the indices $\alpha_L$
and $\alpha_R$ of luminosity and radius functions were calculated. For each
choice of spatial model and luminosity and radius function 600 populations were
simulated and reconstructed. We performed these tests for all
combinations of the indices $\alpha_L = -3, -2, -1$ and $\alpha_R = -2, -1, 0$.
For model mSp4 the mean of the reconstructed $\alpha$ values and
their standard deviations are listed exemplarily in
Table~\ref{TAB:MCverification}.
\begin{table*}[t]
\centering
\caption{Mean and standard deviation of the reconstructed $\alpha_L$ and
$\alpha_R$ for the toy models using the spatial distribution of the four-arm
spiral model of the ISM density (mSp4).}
\label{TAB:MCverification}
\begin{tabular}{r|ccc}
\hline
$\alpha_R$\textbackslash $\alpha_L$ & -3 & -2 & -1\\
\hline
-2
 & $\alpha_{L, reco}=-3.03\,\pm 0.38$ & $\alpha_{L, reco}=-2.04\,\pm 0.28$ &
$\alpha_{L, reco}=-0.99\,\pm 0.30$ \\
& $\alpha_{R, reco}=-2.04\,\pm 0.77$ & $\alpha_{R, reco}=-1.92\,\pm 0.60$ &
$\alpha_{R, reco}=-1.96\,\pm 0.49$ \\\hline
-1 & $\alpha_{L, reco}=-2.99\,\pm 0.36$ & $\alpha_{L, reco}=-2.01\,\pm 0.29$ &
$\alpha_{L, reco}=-0.99\,\pm 0.38$ \\
 & $\alpha_{R, reco}=-1.05\,\pm 0.67$ & $\alpha_{R, reco}=-0.96\,\pm 0.57$ &
 $\alpha_{R, reco}=-0.94\,\pm 0.59$ \\\hline
0 & $\alpha_{L, reco}=-3.01\,\pm 0.35$ & $\alpha_{L, reco}=-2.00\,\pm 0.32$ &
$\alpha_{L, reco}=-1.03\,\pm 0.39$ \\
 & $\alpha_{R, reco}=0.01\,\pm 0.65$ & $\alpha_{R, reco}=0.07\,\pm 0.63$ &
 $\alpha_{R, reco}=0.08\,\pm 0.70$ \\
\hline
\end{tabular}
\end{table*}
For any combination, the mean of the reconstructed $\alpha$ agrees with the true
value within $\Delta \alpha < 0.1$. These results are remarkably
consistent between spatial models. In particular, for all models and
combinations of $\alpha_{L, \mathrm{true}}$ and $\alpha_{R, \mathrm{true}}$,
the reconstructed $\alpha_L$ and $\alpha_R$ are always compatible with the true
values of the simulations. An example is given in Fig.~\ref{FIG:MCverification}
showing for all spatial models the distribution of reconstructed
$\alpha_L$ and $\alpha_R$ given true values of $\alpha_{L,\mathrm{true}} = -2$
and $\alpha_{R,\mathrm{true}} = -1$.
The reconstructed values are always centred at their respective true
values, with standard deviations around $0.3$ and $0.6$. We repeated this exercise
for varying values to investigate
the influence of the boundaries set on the luminosity and radius. No effect on the reconstructed $\alpha_L$
and $\alpha_R$ could be observed.\\
In Fig.~\ref{FIG:mSp4_hist}, we show one-dimensional luminosity and radius
distributions, which are derived from the 600 samples drawn from
model mSp4 for $\alpha_{L,\mathrm{true}} = -2$ and $\alpha_{R,\mathrm{true}} =
-1$.
The blue line indicates the bin-wise mean of the luminosity and radius
distribution for the whole population averaged over the 600
simulated samples, respectively. The blue shaded regions depicts the standard deviation
accordingly. Likewise, the orange area shows the distribution for the selected samples,
each comprised of 16 extended sources within the sensitivity range of the HGPS
and with known $L$ and $R$. Besides the large spread of the latter
distribution it is obvious that the distribution of detected sources on
average does not resemble the global distribution. Therefore, it is inevitable 
for the reconstruction to properly account for the observation bias.
The result of our reconstruction method is indicated in green. The solid line
shows a power law with the mean reconstructed index. The power law is normed to the
total number of sources of the simulated populations and perfectly matches the
input distribution. The green shaded area represents the bin-wise quartile deviation.

\subsubsection{Result}
The derived power-law indices of $P(L,R)$ for the parent source population
of the HGPS sample under the assumption of different underlying spatial source
distributions are listed in Table~\ref{TAB:fit}. Again the results are fairly
consistent among the spatial distributions with average values of
$\left<\alpha_L\right>=-1.77$ and $\left<\alpha_R\right>=-1.26$. 
Owing to the small sample size used in this study the errors on these
reconstructed values are expected to be comparable to those listed in Fig.~\ref{FIG:MCverification}
($\Delta\alpha_{L}\sim0.3$, $\Delta\alpha_{R}\sim0.6$). Besides this statistical
uncertainty, an additional cause of error is the choice of the boundaries of L
and R. In general, deviations of the results due to this choice are found to be
less than the stated errors, while the upper bound of the luminosity affects the
reconstructed values most. However, this upper bound is well constrained since,
according to the HGPS sensitivity, the chance of detecting a source with $L\geq
10^{34.8}\,\mathrm{ph\,s^{-1}}$ (disregarding the availability of a distance estimate) is close to one.
In comparison, previous studies on the VHE source population not taking into account source sizes or the
inhomogeneity of the sensitivity, have estimated the luminosity function to
follow a power law with harder index (e.g. $-1>\alpha>-1.5$; \cite{CasanovaDingus}).
\begin{table}[htb]
\centering
\caption{Reconstructed power-law indices of the joint PDF $P(L,R)$ for different spatial distributions based on:
a) SNRs (mSNR), b) PWNe (mPWN), 
c) free electrons (mFE), d) four-spiral arm model of the ISM density (mSp4), and
e) two-spiral arm model of the ISM density with Galactic bar (mSp2B).}
\label{TAB:fit}
\begin{tabular}{lcccccc}
\hline
Model & $\alpha_{L}$ & $\alpha_{R}$
\\
\hline
mSNR & -1.70 & -1.19 \\
mPWN & -1.81 & -1.13 \\
mFE & -1.94 & -1.21 \\
mSp4 & -1.64 & -1.17 \\
mSp2B & -1.78 & -1.62 \\
\hline
\end{tabular}
\end{table}

\section{Comparison with observable quantities}
\label{SEC:validation}
In order to probe the validity of the derived models, for each model we
compared the distribution of observable quantities from simulated source
populations, namely Galactic longitude and latitude, fluxes, and angular
extents, with those from observations. For each model a set $M$ of
$3000$ synthetic source populations $S \in M$ were simulated. Because the
distribution functions for source positions and source properties were
fixed, we were able to estimate the total number of sources in the population
based solely on the number of observed sources. Thus, it was not necessary to
limit the analysed sample to extended sources, but we could increase the accuracy of our prediction by determining the expected number of detectable sources according to
Eq.~\ref{EQ:scale_threshold}, including point-like sources, that match
the 64 sources in the HGPS that pass this criterion. The numbers of sources per
population for the individual models are listed in Table~\ref{TAB:properties}
and further discussed in Sec.~\ref{SEC:total_number}. The distributions that are
investigated in this section are solely derived from those detectable sources
within the sensitivity range of the HGPS $S_{det} \subset S$ according to
Eq.~\ref{EQ:scale_threshold}.

\subsection{Flux and angular extent}
\label{SEC:flux_extent}
\begin{figure}
        \centering
                \includegraphics[width=0.45\textwidth]{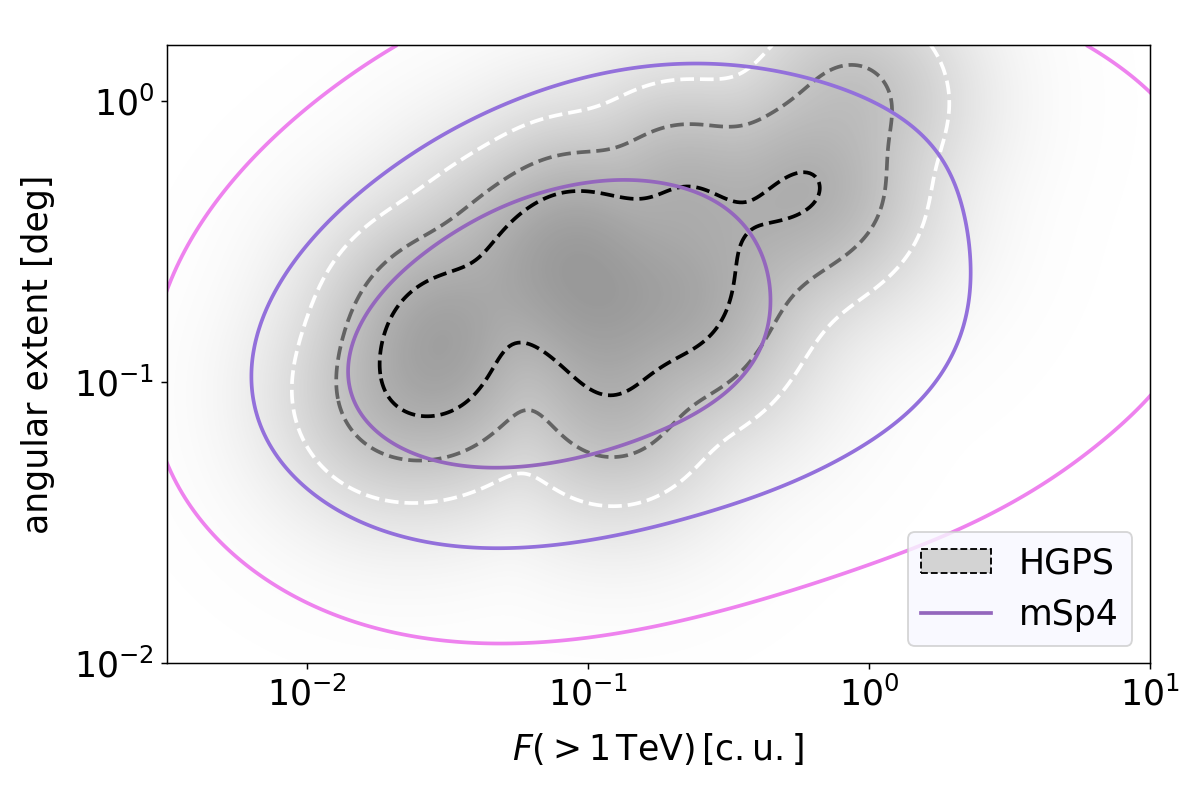}
        \caption{Kernel density estimation for the flux-extent distribution of the
        HGPS sample (grey) and mSp4 (orange). The contour lines indicate
        the $1\sigma$, $2\sigma,$ and $3\sigma$ containment fraction. Flux is given in
        `crab units' (c.u.)}
        \label{FIG:kde_contour}
\end{figure}

The observed flux and angular extent of a source both depend on the
distance of the source to the observer. In addition to this observational
correlation, an intrinsic correlation between the luminosity and
radius of sources can shape the observed distribution of fluxes and angular
extents. Instrumental selection effects, such as the dependency of the
sensitivity on the angular extent, can shape this distribution as well. To
account for correlations, we compared the observed two-dimensional flux-extent
distribution by means of its PDFs against the model prediction. We derived the
PDFs from kernel density estimations for which
the optimal bandwidth for a Gaussian kernel was derived individually in the range $[10^{-3},
10]$ via the \textit{GridSearchCV} method from the Python package scikit-learn. The distributions are very similar
among the models, thus only the result for mSp4 is shown in comparison
with the HGPS distribution in Fig.~\ref{FIG:kde_contour}.
Contour lines indicate the $1\sigma$, $2\sigma$, and $3\sigma$ containment
fraction of the derived PDFs for the observed distribution and the
predicted distribution. For all distributions we observe an increase of the
angular extent with flux, although this correlation
appears less pronounced for the model distributions. Additionally, model distributions are noticeably wider
than the observed distribution. This discrepancy is further reflected in the
fraction of extended sources in the sample of detected sources. While for the
HGPS we yield a fraction of $78\,\%$ (50 out of 64) of extended sources, for the models we observe
on average a fraction between $23\,\%$ - $38\,\%$, the rest being point-like. This discrepancy might 
be an effect that is intrinsic to the HGPS. More point sources might be present in the data set but are ``lost'' 
in extended sources owing to source confusion \citep{SourceConfusion} or detected and later merged with 
an overlapping source. We did not account for these effects in our model.
Alternatively, the model assumptions might not reflect reality and the number of
point-like sources is overestimated: this might be an effect of the simplified
definition of source extent, namely describing distinct and complex morphologies
altogether by a single parameter. In addition, it is likely that the assumed
independence of radius and luminosity and the power law for the source radius do
not capture the true nature of this source property. Given the connection to the
observation bias and especially the impact on the detectability of nearby
sources, this relation is worth investigating in a follow-up study.
Nevertheless, the flux-extent distribution of our models and observations are
considered to be close enough to make reasonable predictions in the following.

\subsection{Spatial distribution of sources}
\begin{figure}[htb]
        \centering
                \includegraphics[width=0.45\textwidth]{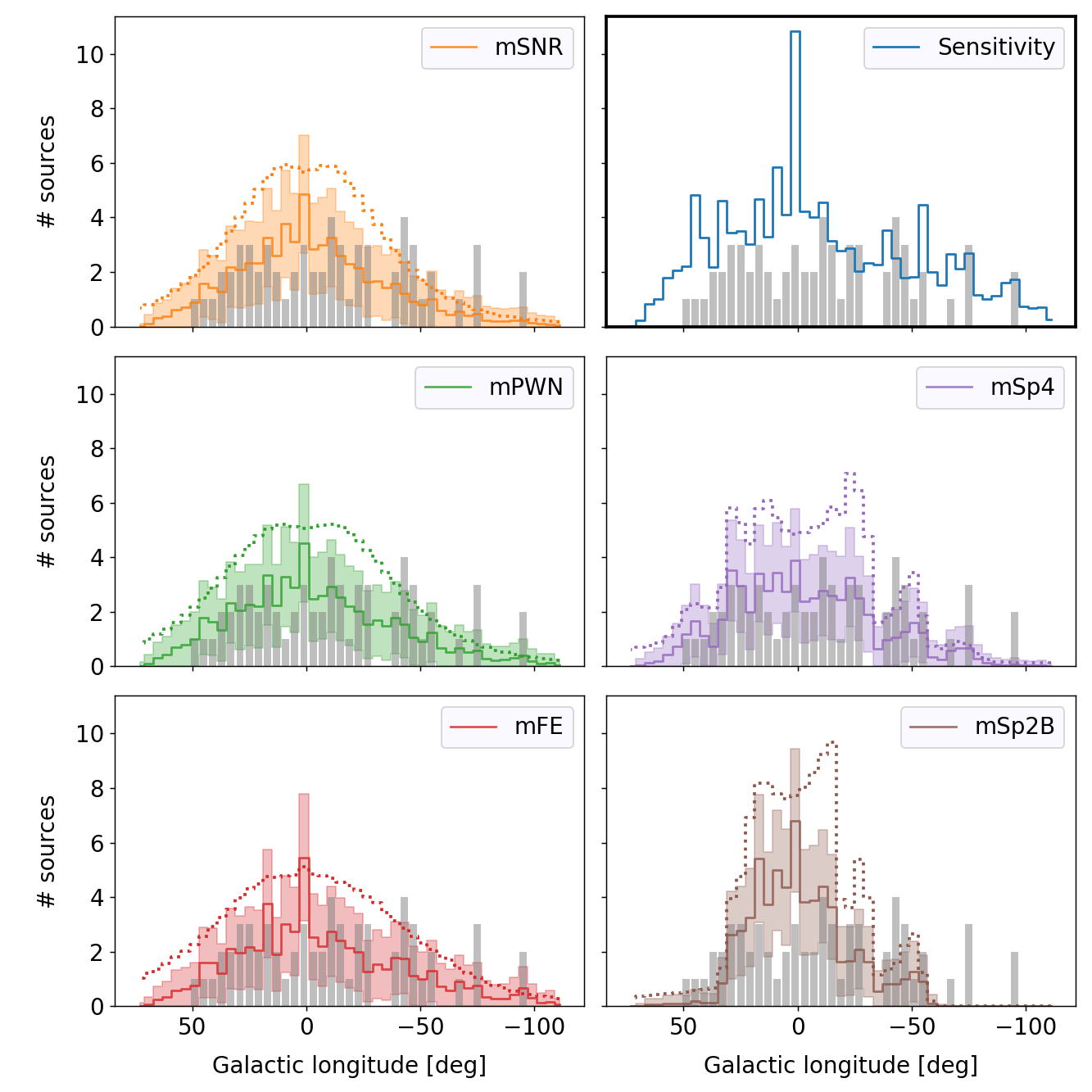}
        \caption{Comparison of the longitudinal source distribution between HGPS
        sources (grey bars) and predictions of different models (coloured lines).
        The dotted lines represent the modelled source densities within the HGPS region,
        arbitrarily scaled for the sake of visibility. In addition, the dynamic range
        of the HGPS sensitivity is shown on the top right panel. Details are given in
        the text.}
        \label{FIG:glong_dist}
\end{figure}

\begin{figure}[htb]
        \centering
                \includegraphics[width=0.45\textwidth]{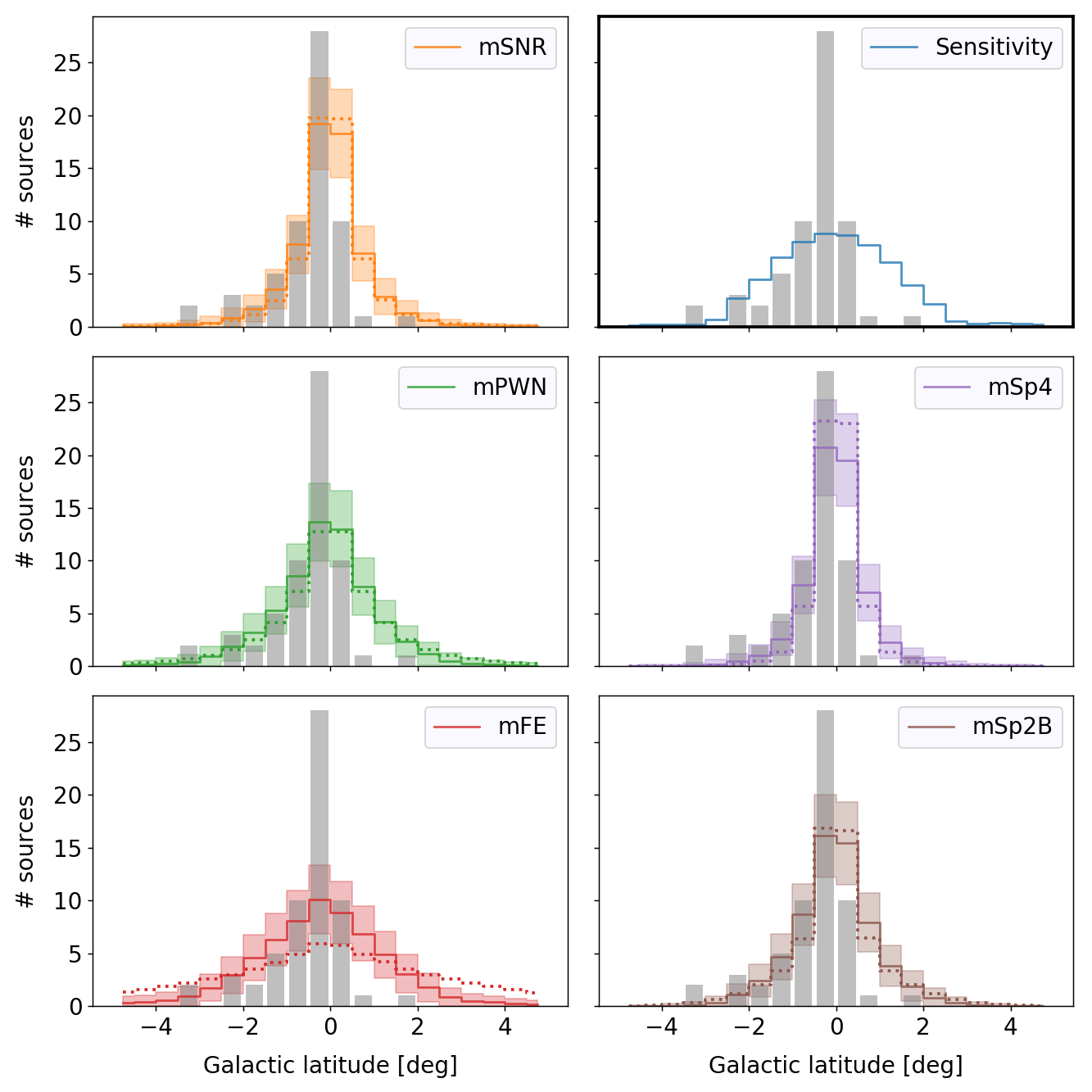}
        \caption{Comparison of the latitudinal source distribution as in
        Fig.~\ref{FIG:glong_dist}}
        \label{FIG:glat_dist}
\end{figure}

The source distributions in Galactic longitude for the probed models are
depicted in Fig.~\ref{FIG:glong_dist}.
The shaded region shows the standard deviation around the mean value of the
different samples $S_{det}$ while the solid line represents the mean. For comparison,
the observed distribution of the HGPS is given by the grey bars. Furthermore,
the source density of the respective model within the FOV of the HGPS
is shown by the dotted line, which is scaled for better visibility
($\int_{FoV}\rho(\mathbf{x})\,d\mathbf{x} = 2 N_{HGPS}$).
On the top right panel of Fig.~\ref{FIG:glong_dist} the dynamic range of the
sensitivity over Galactic longitude is shown, which is expressed by
$F_{min}^{-3/2}$, where $F_{min}$ is the point-source sensitivity of the HGPS
map at $b=0^{\circ}$ and the corresponding longitude bin. Here, for 
a given luminosity $F_{min}^{-3/2}$ is proportional to the sampled volume.
For better visibility this distribution is scaled in the same way as the source densities.\\
The models are generally in good agreement with
observations, although for mSp2B the longitude distribution appears to be
somewhat too narrow as it falls off too steeply in the outskirts of the Galactic
plane. The result for mSp2B suggests that if the ISM density distribution, which
is used as proxy for the distribution of regions with high star formation rates,
indeed follows the assumed shape, the VHE source population must comprise at
least one source class that evolves outside those regions.
In the central regions all models commonly tend to overpredict the actual source distribution.
According to the models, this is the region of highest source density.
Therefore, we can arguably expect that the detection of sources is also affected by source
confusion and an increased background resulting from the
existence of bright diffuse emission in the Galactic ridge region
\citep{GCDiffuse}; both effects lead to a deficit of detected sources in
that region. A future iteration of the model, taking both
effects into account, will be required to unambiguously test whether this discrepancy in the
central region can be attributed to an inaccurate spatial source
distribution. Regarding the distribution of the source density we observe that,
for model mSp4, peaks in this distribution align well with peaks in the observed
source distribution. This might be suggestive that the Galactic population of
VHE $\gamma$-ray sources indeed follows a similar spiral structure as derived from ISM measurements.
However, with the inhomogeneity of the sensitivity, which yields similar distributions for
detectable sources independently of the probed source distribution model, it is not
feasible to make strong claims. Quantitatively we investigate the compatibility
between model predictions and observations by means of the
Kolmogorov-Smirnov test statistic $d_{n}$ as follows:
\begin{equation}
d_{n} = \sup_{x}\,\lvert F_{n}(x)-F_{0}(x)\rvert
,\end{equation}
where $F(x)$ is the cumulative distribution over the variable $x$
(i.e. Galactic longitude). In this equation, $F_{0}(x)$ is derived from the mean distribution
of a given model as shown by the solid line in Fig.~\ref{FIG:glong_dist}.
For each simulated population $d_{n}$ is calculated with the cumulative
distribution $F_{n}(x)$ of detectable sources, yielding the probability
distribution $P(d_{n})$. From the cumulative distribution of observed sources we
derive $d_{HGPS}$ and calculate the p-value $P(d_{n}\geq d_{HGPS})$. The values
listed in Table~\ref{TAB:pvalue} confirm that only the longitude
distribution of model mSp2B is incompatible with observations at a
level of significance of $5\,\%$.\\
\begin{table}[htb]
\centering
\caption{Compatibility of the modelled longitude and latitude distribution with
observations}
\label{TAB:pvalue}
\begin{tabular}{ccccccccc}
\hline
 & \multicolumn{2}{c}{p-value}\\
\cline{2-3}
 &  longitude & latitude \\
\hline
mSNR & 0.27 & 0.73 \\
mPWN & 0.52 & 0.36 \\
mFe & 0.77 & 0.08 \\
mSp4 & 0.25 & 0.88 \\
mSp2B & 0.03 & 0.40 \\
\hline
\end{tabular}
\end{table}
Source distributions in Galactic latitude are presented in
Fig.~\ref{FIG:glat_dist} in the same way as for the longitudinal distributions.
While for the latitude distribution all models are compatible with observations
according to the statistical test, we can see some obvious deviation
in this plot. The number of observed sources is falling rapidly outwards from the
Galactic disc. The most prominent feature of the observed distribution is an
asymmetry towards the southern sky that is not covered by any of the assumed
models. Given that here we average over a broad source distribution over
Galactic longitude, source confusion can be assumed to play a less pronounced
role and the simulations correctly reflect the observation bias. Thus, this
asymmetry appears to be a real feature of the spatial source distribution ({\it
cf.} e.g. \cite{WarpedMilkyWay}), which is not accounted for in the models.
Besides this, data show a stronger peak of the latitude distribution towards the
Galactic equator compared to simulations. This is most notably the case for the
model mFE, whose flatter distribution appears to be in tension with observation.

\section{Global properties of the Galactic source population}
\label{SEC:prediction}
While in the previous section it is shown that most models can describe
observations reasonably well within the sensitivity range of the HGPS, in the
following section these models are used to predict global properties of the
Galactic VHE source population, namely the total number of sources in the Milky
Way, their contribution to the observed $\gamma$-ray flux, and their cumulative luminosity.

\subsection{Total number of sources}
\label{SEC:total_number}
\begin{figure}[htb]
        \centering
                \includegraphics[width=0.45\textwidth]{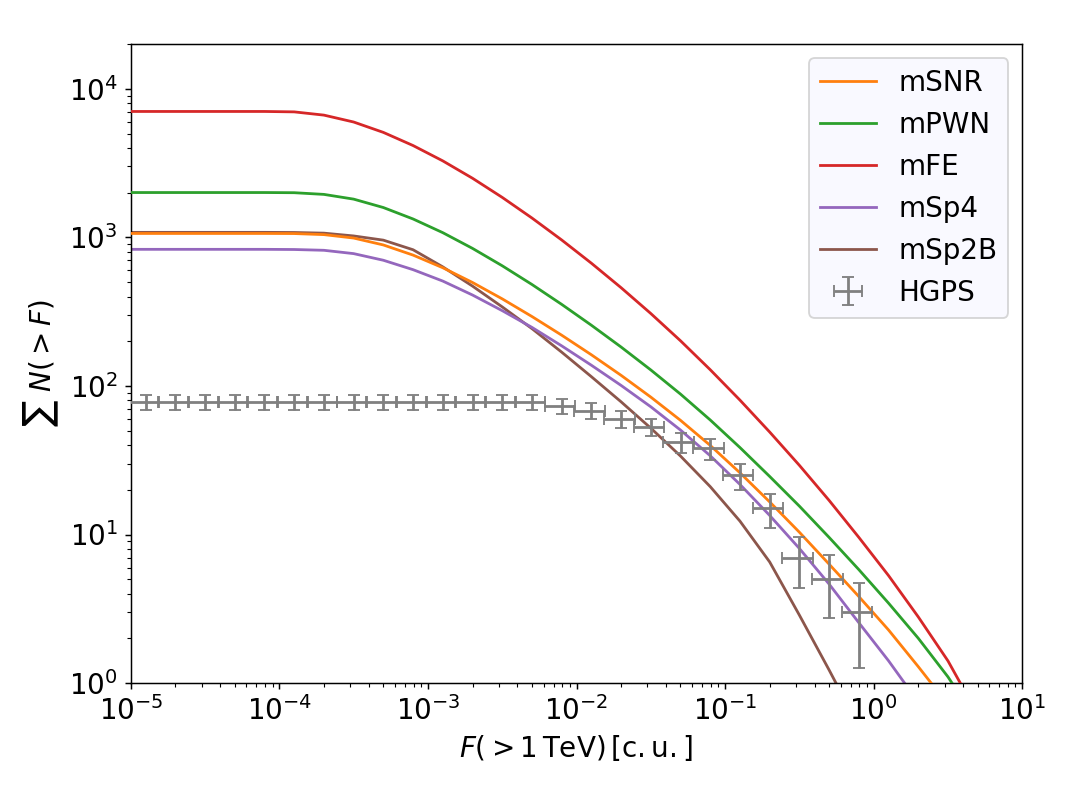}
        \caption{Cumulative source distribution over flux (log(N)-log(F)). The HGPS
        source distribution is given by grey points with horizontal error bars
        depicting the bin width and vertical error bars showing the Poissonian error.
        Coloured lines represent the distribution of the whole Galactic source population
        averaged over the different
        realisations of the respective model. As the whole population also includes 
        sources outside the FoV or too extended to be detectable, most models also overshoot the HGPS
        data in the range of completeness for point-like sources. Only mSp2B is in clear 
        conflict with HGPS data.}
        \label{FIG:logn-logs}
\end{figure}

\begin{figure}[htb]
        \centering
                \includegraphics[width=0.45\textwidth]{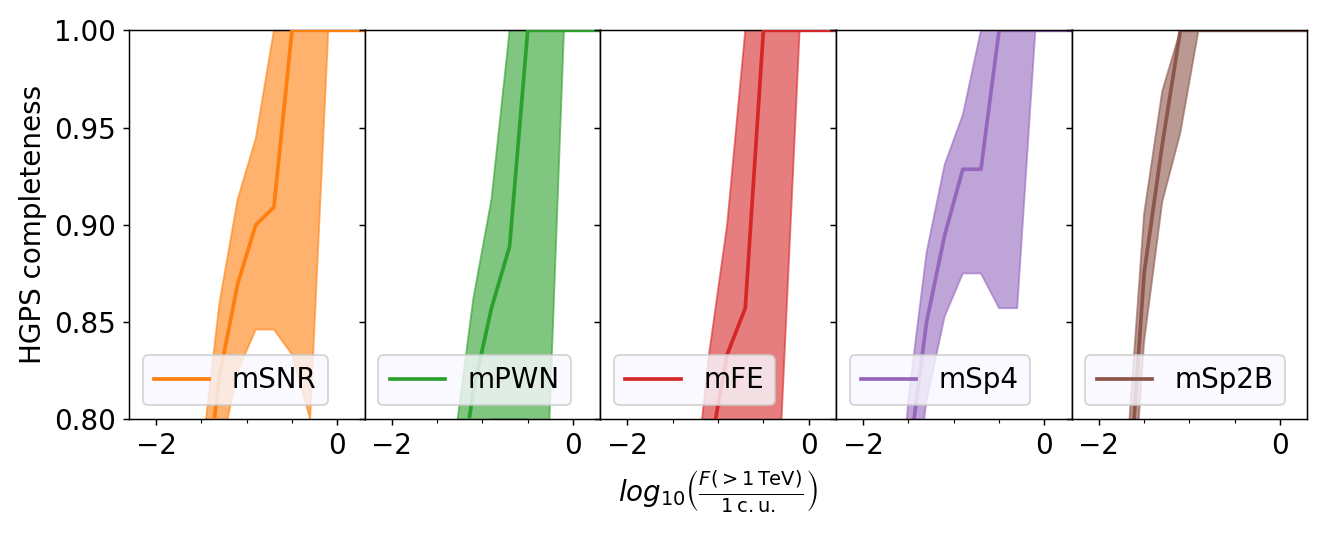}
        \caption{Fraction of detectable sources in the FOV of the HGPS
        cumulatively over source flux. The shaded regions represent the interquartile range; the
        lines depict the median.}
        \label{FIG:completeness}
\end{figure}

As described in Section~\ref{SEC:validation}, we can derive the average number of detectable
sources $\left<N_{det}\right>$ according to Eq.~\ref{EQ:scale_threshold} for any
given total number of sources in a population $N_{tot}$:
$\left<N_{det}(N_{tot}) \right>$. With the probability of detecting 64 HGPS sources given
by the Poissonian $P_{\left<N_{det}\right>}(64)$, the number of Galactic sources
was derived from the maximum of the distribution
$f\left(N_{tot}\right)=P_{\left<N_{det}\left(N_{tot}\right)\right>}(64)$ and
cited errors from the corresponding $68\,\%$ containment area around this
maximum. These numbers vary considerably among the probed models, ranging from
831 sources (mSp4) up to 7038 sources (mFE) (see Table~\ref{TAB:properties}).
\begin{table}[htb]
\centering
\caption{Population properties according to the probed models. The distribution
of the total number of Galactic VHE $\gamma$-ray sources $N$ and their combined
luminosity $L$ and flux $F$. The luminosities are characterised
by the mean and standard deviation as this quantity is almost symmetrically
distributed over the different realisations. In contrast, the median and
quartile deviation are used as a more robust description of the total flux $F$,
whose distribution is strongly affected by outliers that stem from (rare) nearby
sources.}
\label{TAB:properties}
\begin{tabular}{lccc}
\hline
Model & $N$ & $L$ [ph$\,\mathrm{s^{-1}}$] & $F$ [ph$\,\mathrm{cm^{-2}\,s^{-1}}$] \\
 & & Mean / Std & Median / QD \\
\hline
mSNR  & $1063_{-126}^{+137}$   & (1.73 / 0.16)$\,\cdot 10^{36}$ & (4.89 /
1.41)$\,\cdot 10^{-10}$
\\
mPWN  & $2004_{-239}^{+259}$   & (2.47 / 0.18)$\,\cdot 10^{36}$ & (7.74 /
1.86)$\,\cdot 10^{-10}$
\\
mFE   & $7038_{-839}^{+912}$ & (6.32 / 0.26)$\,\cdot 10^{36}$ & (1.54 /
0.16)$\,\cdot 10^{-9}$
\\
mSp4  & $831_{-98}^{+107}$   & (1.59 / 0.17)$\,\cdot 10^{36}$ & (3.38 /
0.70)$\,\cdot 10^{-10}$
\\
mSp2B & $1081_{-128}^{+140}$     & (1.44 / 0.14)$\,\cdot 10^{36}$ & (1.96 /
0.20)$\,\cdot 10^{-10}$
\\
\hline
\end{tabular}
\end{table}
Although sources are treated generically as VHE $\gamma$-ray emitters in this
model, that is no source type is explicitly assumed, a source count as
high as seen for model mFE is challenging for the paradigm that SNRs and PWNe
are the dominant source classes of VHE $\gamma$-ray emission. With a Galactic
supernova rate of one per $40\,\mathrm{yr}$~\citep{Tammann1994} a source count
of 7000 implies a maximum age of emitters of $\sim 3 \cdot 10^{5}\mathrm{yr}$.
Interestingly, the models mSNR, mSp4, and mSp2B yield very similar results regarding
the total number of sources. The similarity between model mSNR and mSp4 is also
apparent in the cumulative source distribution over flux (log(N)-log(F)) as
shown in Fig.~\ref{FIG:logn-logs}, while mSp2B shows distinct differences.
In Fig.~\ref{FIG:logn-logs} the distribution of observed sources
in the HGPS is given by grey points with Poissonian errors. The mean distribution for the whole
Galactic population\footnote{That includes non-detectable sources and sources
outside the FOV of the HGPS.} according to the different models is given by coloured lines.
It appears that model mSp2B on average does not comply with the
observed distribution, especially for $F>0.03\,\mathrm{c.u}$ (c.u.: integral
flux of the Crab Nebula above $1\,\mathrm{TeV}$) yielding too few sources in this regime. In
contrast, the other four models allow for sources of high flux (e.g. > 0.1 c.u.)
being undetected by the HGPS.
The latter point is more clearly shown in Fig.~\ref{FIG:completeness}.
In this figure the predicted completeness range of the HGPS is shown; more
precisely, the figure shows the median fraction of detected sources within the FOV for
sources exceeding a given flux level. This number decreases with
decreasing flux levels as fainter sources are less likely to be detected. For
most models a kink in the distribution is observed between
$0.1\,\mathrm{c.u.}-1\,\mathrm{c.u.}$. That is caused by the limited sensitivity
of imaging atmospheric Cherenkov telescopes (IACTs) to extended sources. 
For flat spatial source distributions the
likelihood for close-by and, therefore, bright and extended sources to be found in the Galactic
population increases. For model mSp4 we find that $2\pm1$ and for model mPWN
$3\pm2$ sources exceeding the threshold of $1^{\circ}$ are expected to be
found in the Galactic population. To probe this regime either different data analysis techniques or
different observation techniques, for example with water Cherenkov telescopes that can make use
of a large FOV, can be exploited.
Indeed, two extended sources, Geminga~\citep{milagro, hawc} and 2HWC J0700+143~\citep{hawc},
have been detected this way, which is in good agreement with both predictions.
\\Taking this one step further, we used these models to predict the number of sources
that are detectable with the next generation of IACTs, CTA. Aiming for a
point-source sensitivity of $2\,\mathrm{mCrab}$ in the longitude range $|l|<60^{\circ}$ and latitude range
$|b|<2^{\circ}$~\citep{cta_science}, the predicted numbers of detectable sources
lie in the range 295 (mSp4) - 457 (mFE). Since most of these sources are expected
to appear point-like to CTA this number does not suffer considerably from a
degradation of the sensitivity with source extent. Especially, this estimation
is not affected by the inaccuracy of the description of source radii inherent to
our models. The derived number is valid for the boundaries chosen
for the luminosity. The implications for probing a larger dynamical range as it
might be possible with CTA are discussed in Sec.~\ref{SEC:conclusion}. With the
HGPS providing 53 sources in the region to be observed by the CTA Galactic plane scan, 
the CTA sample would increase the
current source sample substantially by a factor between 5 - 9 according to
these models.

\subsection{Flux of unresolved sources}
\begin{figure}
        \centering
                \includegraphics[width=0.45\textwidth]{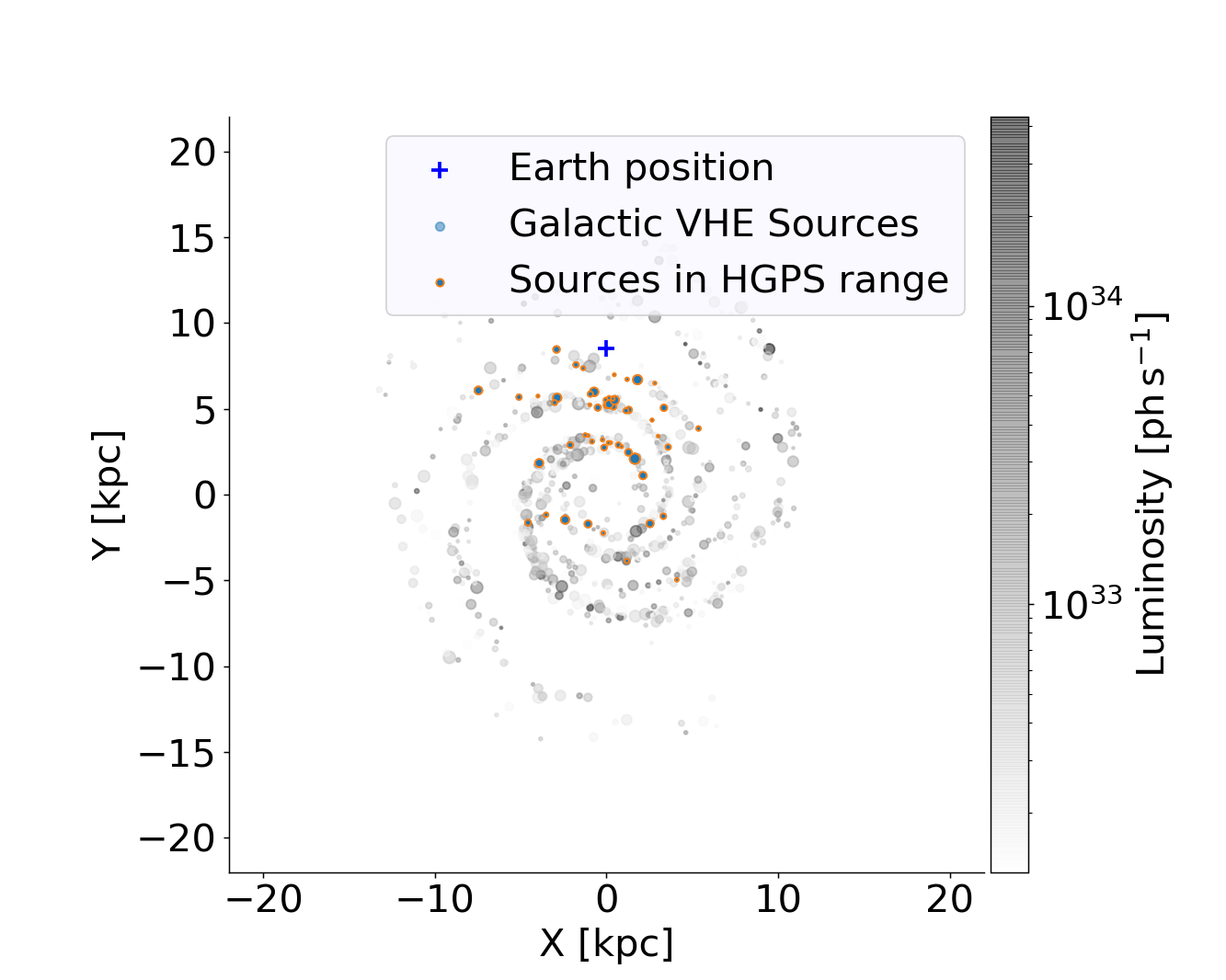}
        \caption{Example realisation of a synthetic population of VHE $\gamma$-ray
        sources for model mSp4. Source luminosities are given by the colour scale;
        source radii are proportional to circle radii.
        Detectable sources within the HGPS sensitivity are indicated with orange circles.}
        \label{FIG:skymap_faceon}
\end{figure}

Although the HGPS is expected to comprise only $1\,\%-9\,\%$ of all sources
in the Milky Way, these sources can already account for a significant fraction
of the measurable flux (total flux given in Table~\ref{TAB:properties}).
In Fig.~\ref{FIG:skymap_faceon} we give an illustrative example of one realisation of a synthetic VHE
source population for model mSp4 in a face-on view of the Galaxy. For all
of the 831 sources the luminosity and radius are encoded in the colour and
radius of the circles representing them (radius not to scale), while sources that
can be detected with the HGPS sensitivity are additionally denoted by an orange
circle.
\begin{figure*}
        \centering
                \includegraphics[width=0.95\textwidth]{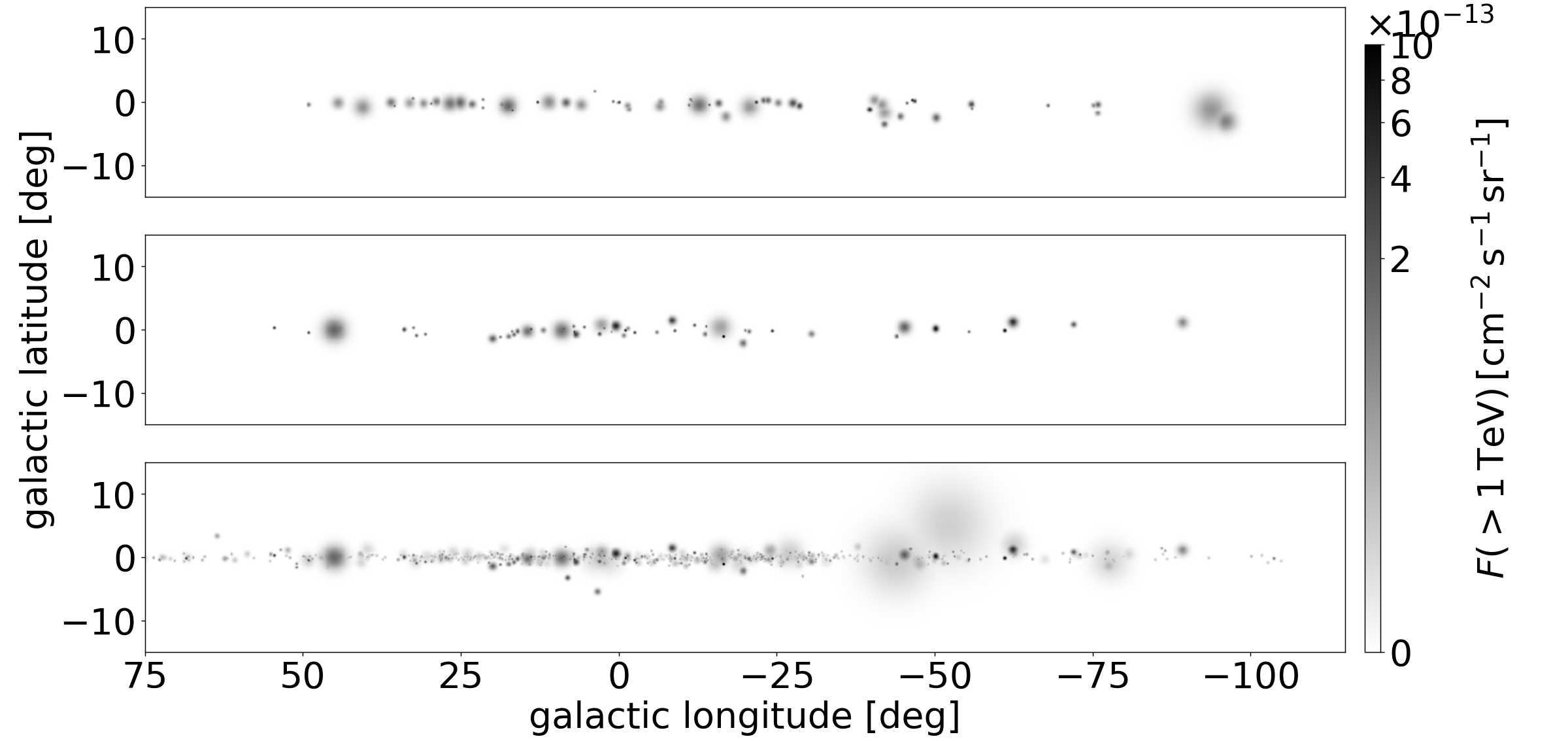}
        \caption{Sky maps of the VHE source population in the HGPS region. Top: From
        the HGPS catalogue. This map has been constructed by assuming a spherical
        source shape for better comparability with the simulations. Middle: Detectable
        sources from the synthetic population shown in Fig.~\ref{FIG:skymap_faceon},
        based on the mSp4 model. Bottom: All sources from this simulated realisation.}
        \label{FIG:skymap}
\end{figure*}
Corresponding sky maps of the fluxes for this realisation of the population are shown in Fig.~\ref{FIG:skymap}.
The middle panel of Fig.~\ref{FIG:skymap} shows the sample of detectable sources.
In comparison with the HGPS sample shown in the top panel a similarity of the two
samples is recognisable, although the HGPS clearly shows a larger fraction of
extended sources. The discrepancy in the ratio of extended to point-like sources
between observation and model prediction was already discussed in
Sec.~\ref{SEC:flux_extent}. The sky map of the same synthetic population when
observed with infinite sensitivity is shown in the bottom panel of Fig~\ref{FIG:skymap}.
There, a band of faint sources along the Galactic plane, which can contribute to the
unresolved, large-scale VHE $\gamma$-ray flux, is clearly seen.
Focussing only on the region scanned by the HGPS, the flux of all sources detected by
the HGPS exceeds the prediction of model mSp2B by $\sim24\,\%$ as already
indicated by the log(N)-log(F) distribution (see Fig.~\ref{FIG:logn-logs}). The
other four models predict that unresolved sources make up about
$13\,\%-32\,\%$ of the total flux stemming from the source population within
this region.
The H.E.S.S. measurement of large-scale diffuse emission in the HGPS from regions
that do not contain any significant $\gamma$-ray emission quotes a similar
number of $\sim28\,\%$ of the total measured VHE emission in large-scale diffuse
emission~\citep{diffuse}. However, these numbers are not directly comparable
since the sky regions they are derived from are not identical. While the model
allows us to remove detectable sources easily, complex exclusion regions were applied for the H.E.S.S. measurement
 to exclude contribution from sources,
most of which accumulate at small Galactic latitude values. Still, these numbers
are suggestive that unresolved sources might very well be the dominant component
of the measured diffuse emission.
\\Additionally, the two prominent extended sources that are seen at $\sim
-50^{\circ}$ longitude on the sky map at the bottom of Fig.~\ref{FIG:skymap}, but
not in the sky map in the centre, demonstrate the effect of the maximum extent
detectable by H.E.S.S., which has been discussed with respect to the catalogue
completeness (see Sec.~\ref{SEC:total_number}).
\subsection{Luminosity of the Galactic source population}
\label{SEC:lum}
Using the mean photon energy of
$3.52\,\mathrm{erg\,ph^{-1}}$ from Sec.~\ref{SEC:lum_size}, 
the total luminosity of the Galactic VHE source population is estimated to lie
in the range
$(5.07\cdot10^{36}-2.22\cdot10^{37})\,\mathrm{erg\,s^{-1}}$.
Assuming that $\gamma$-ray sources are the dominant contribution to the overall
VHE $\gamma$-ray luminosity of the Milky Way
and that the diffuse emission originating from propagating cosmic rays adds only
a small contribution, these values can be compared with the total luminosity at
megaelectronvolt and gigaelectronvolt energies
($\sim3\cdot10^{38}\,\mathrm{erg\,s^{-1}}$ and
$\sim8\cdot10^{38}\,\mathrm{erg\,s^{-1}}$, respectively; \cite{energy_budget}).
The $\gamma$-ray luminosity of the Milky Way at VHE turns out to be one to two
orders of magnitude lower than in those two lower energy bands. This
demonstrates that the presented models are compatible within the available energy budget constraints
and indicates a drop in luminosity between the HE and VHE ranges.

\section{Conclusions}
\label{SEC:conclusion}
We present models of the VHE $\gamma$-ray source population of the Milky
Way, based on different assumptions of the spatial source distributions. 
Power-law indices of luminosity and radius
functions of the population are derived from a subsample of the HGPS source catalogue 
(namely, extended sources with known distances) and its
sensitivity. We pay special attention to correction of the observation bias.
The validation of this bias correction is done with simulated toy models and demonstrates
very good reconstruction capabilities. Furthermore, the simulations demonstrate that relying 
on the detected set of sources with no bias correction  gives more or less arbitrary results.
In this context it has to be noted that the limitation to the range of completeness does not
completely avoid this problem because the completeness relates to point-like sources and does not
apply to sources of larger extension.\\
A comparison of the source models with HGPS observations demonstrates a reasonable agreement. 
Despite a lack of asymmetry in the latitude distribution seen in
all models as compared to the observed distribution (see
Fig.~\ref{FIG:glat_dist}), simulations approximately reproduce the spatial HGPS
source distribution as well as the distribution of source fluxes and extents.
Only the model mSp2B is disfavoured owing to its distinctively different
longitude distribution in the HGPS sensitivity range and the clear
under-prediction of the total flux from VHE $\gamma$-ray sources. 
All models under-predict the fraction of extended sources in the detectable sample, which can
be attributed to either effects in the construction of the HGPS catalog (e.g. source confusion)
or shortcomings of the model (e.g. invalidness of underlying assumptions) or a combination of both.\\
Despite the rather limited statistics in the sample of HGPS sources, our
data-driven approach, which minimises the degrees of freedom, allows for
relatively good predictions to be made. The derived models can be used to study possibilities and limitations of VHE observations. Examples are expectations for future instruments,
an assessment of the amount of yet unresolved sources in a large-scale diffuse
emission measurement depending on the sensitivity threshold, and the study of
observational challenges like source confusion. We derived
the total number of VHE sources in the Milky Way and the total luminosity and flux of the Milky Way in VHE $\gamma$-rays.
Disregarding the one disfavoured model, of the four remaining
models, mPWN, mSNR, and mSp4 do not deviate substantially from one another.
While model mFE also yields similar results regarding the distribution of
source properties, the predicted number of sources within the Milky Way exceeds
the predictions of the other models by a factor $>3$. Thus, the predicted range
of VHE sources in our Galaxy is 800 to 7000. The scatter in the total luminosity
and the total flux lies with a factor $\sim 4$ in the range of $(1.6-6.3) \cdot
10^{36}$~ph~s$^{-1}$ and $(3-15) \cdot 10^{-10}$~ph~cm$^{-2}$~s$^{-1}$,
respectively. A significant fraction, $(13-32)\%$, of the $\gamma$-ray emission
of sources within the HGPS region is attributed to yet unresolved sources and contributes to the measured diffuse emission.
With a foreseen sensitivity of $2\,\mathrm{mCrab}$ in the central Galactic plane, 
CTA should be able to increase the known Galactic VHE $\gamma$-ray source sample
by a factor between 5 - 9.\\
It should be noted that the derived properties depend not only on the chosen spatial distribution model but also
on the luminosity range covered by the model. Especially, the low-luminosity limit
can have a significant impact on the expected number of sources and the energy
budget. The choice of this limit was made to encompass the HGPS data, and
it is possible that weaker sources (or a subdominant class of less luminous sources) 
can be on the verge of detection and not being
identified with the present, inhomogeneous sensitivity. Expanding the dynamical
range towards lower luminosities in the models increases the fraction of sources that lie outside the
sensitivity limit and thus increases the expected number of VHE sources in the
Galaxy and simultaneously increases (although to a lesser degree) predictions for fluxes and luminosities.
Another factor that affects the estimation of the size of the
population is source confusion. Given the predicted number of sources in the
Galaxy, confusion of sources in the FOV and their ascribed fluxes
seems to be inevitable.
If source confusion already underlies the HGPS detected data set, the true size of
the VHE source population can be expected to be even larger and luminosity and radius functions are
subject to an additional bias. \\
Consequently, improvements of these models planned for the future 
include the correct treatment of source confusion in the comparison between simulations and data. 
Additionally, inclusion and proper treatment of sources with incomplete information (point-like sources/no 
known distance estimation) can increase the statistics of the sample.
The incorporation of spectral information will allow for 
comparisons with neighbouring energy bands, making use of the wealth of
information at Fermi-LAT energies and the complementarity of wide FOV HAWC observations.\\
In order to study the physics of VHE $\gamma$-ray sources, differentiation is
required between the various source classes, such as SNRs and PWNe as the likely
dominating contributors, and smaller contributions from other classes. This goes
in conjunction with a widening of the parameter space that is needed to include
physical source properties rather than the phenomenological observables used
in this work. Although weakly constrained with the currently available data,
this inclusion of physical source modelling for the characterisation of the
population properties of VHE source classes is the logical next step after
the in-depth studies of single objects and the description of the detected
samples of sources so far performed with HGPS data (see e.g. \cite{Cristofari1, Cristofari2}). The presented models can
serve as a basis for such studies. With future surveys of more
sensitive instruments such as CTA the amount of available data will increase, 
allowing for more accurate characterisations of the SNR and PWN populations.
\\

\begin{acknowledgements} 
We gratefully acknowledge fruitful discussions and feedback by Olaf Reimer, Ralf
Kissmann, and Julia Thaler. For a very useful and constructive report which
helped to improve the paper we would like to thank the anonymous referee. This
work was conducted in the context of the CTA Consortium. This paper has gone
through internal review by the CTA Consortium for which we like to thank
Matthieu Renaud and Ullrich Schwanke.
\end{acknowledgements}
\bibliographystyle{aa} 
\bibliography{paper} 
\end{document}